\documentclass[twocolumn,english,aip,jap,reprint]{revtex4-1}
\usepackage[T1]{fontenc}
\usepackage[latin9]{inputenc}
\usepackage{color}
\definecolor{document_fontcolor}{rgb}{0, 0, 0}
\color{document_fontcolor}
\usepackage{babel}
\usepackage{bm}
\usepackage{amstext}
\usepackage{graphicx}
\usepackage[unicode=true,pdfusetitle,
 bookmarks=true,bookmarksnumbered=false,bookmarksopen=false,
 breaklinks=false,pdfborder={0 0 1},backref=false,colorlinks=true]
 {hyperref}
\usepackage{breakurl}

\makeatletter
\@ifundefined{textcolor}{}
{%
 \definecolor{BLACK}{gray}{0}
 \definecolor{WHITE}{gray}{1}
 \definecolor{RED}{rgb}{1,0,0}
 \definecolor{GREEN}{rgb}{0,1,0}
 \definecolor{BLUE}{rgb}{0,0,1}
 \definecolor{CYAN}{cmyk}{1,0,0,0}
 \definecolor{MAGENTA}{cmyk}{0,1,0,0}
 \definecolor{YELLOW}{cmyk}{0,0,1,0}
}

\makeatother

\begin{document}

\title{Ferromagnetic resonance of magnetostatically-coupled shifted chains
of nanoparticles in an oblique magnetic field}

\author{R. Bastardis}

\email{roland.bastardis@univ-perp.fr}

\author{J.-L. Déjardin }

\email{dejardin@univ-perp.fr}

\author{F. Vernay}

\email{francois.vernay@univ-perp.fr}

\author{H. Kachkachi}

\email{hamid.kachkachi@univ-perp.fr}

\affiliation{Laboratoire PROMES CNRS (UPR-8521) \& Université de Perpignan Via
Domitia, Rambla de la thermodynamique, Tecnosud, F-66100 Perpignan,
France}

\date{\today}
\begin{abstract}
We investigate the ferromagnetic resonance characteristics of a magnetic
dimer composed of two shifted parallel chains of iron nanoparticles
coupled by dipolar interactions. The latter are treated beyond the
point-dipole approximation taking into account the finite size and
arbitrary shape of the nano-elements and arbitrary separation. The
resonance frequency is calculated as a function of the amplitude of
the applied magnetic field and the resonance field is computed as
a function of the direction of the applied field, varied both in the
plane of the two chains and perpendicular to it. We highlight a critical
value of the magnetic field which marks a state transition that should
be important in magnetic recording media.
\end{abstract}
\maketitle

\section{introduction}

Organized assemblies of nearly monodisperse nanoparticles are a recent
achievement in materials science which is a result of a long-term
endeavor of many research groups around the world \citep{sun00science,lisieckietal03am,tartajetal03jpd}.
One of the main initial objectives for working towards this goal was
to minimize the effects of volume and anisotropy distributions which
make it more difficult, if not impossible, to access the intrinsic
effects of magnetic nanoparticles. In parallel with this progress
in fabrication and synthesis, several measuring techniques have benefited
from considerable improvements with regards to time and spatial resolution.
Some of the standard techniques, such as ferromagnetic resonance (FMR)
\citep{vonsovskii66pp,gurmel96crcpress}, stage a successful come-back
\citep{duerretal11apl,leeetal11jap,dingetal12apl}. The latter is
a very efficient technique for characterizing assemblies of magnetic
nanoparticles \citep{begin_comm_privee}.

Accordingly, in this work we investigate the FMR characteristics of
a monolayer of chains of (almost) monodisperse \textcolor{black}{Fe}
nanoparticles of $D=20\ {\rm nm}$ in diameter. Within each chain,
the particles are closely packed and touch each other, while the (nearly)
parallel chains are a certain distance from each other, and may be
shifted with respect to each other along their (major) axes. We investigate
the effects of the inter-chain shift and separation on the FMR frequency
and resonance field. These two parameters greatly affect the dipolar
interactions (DI) between the chains, in addition to the size and
shape of the chains. This aspect has been recently mentioned by Varón
\textit{et al}.\citep{Varon_dipolar_scirep2013} in a\textit{ }study
of the dipolar magnetism in ordered and disordered low-dimensional
nanoparticle assemblies. In particular, these authors show that the
DI are no longer negligible in such systems with respect to the usual
prevalence of the exchange coupling in classical materials. In order
to account for the effects of all these parameters, we shall consider
here the more general formalism of DI \citep{beleggiaetal04jmmm272,beleggiaetal04jmmm278,francoetal14jap}
between finite-size magnetic elements of cylindrical shape. From the
applied physics standpoint, this investigation is of interest in the
context of magnetic recording. Indeed, researchers have realized since
the synthesis of FePt nanorods \citep{Wang_nanopillar_ange-chem2007},
that 1D nano-elements have a major advantage compared to more conventional
nanoparticle assemblies. While spherical nanoparticles have magnetic
moments that are difficult to texture, nanochains are usually aligned
with one another due to geometric constraints \citep{Toulemon_these,begin_comm_privee}.
In addition, since a nanochain usually exhibits a magnetic anisotropy
with an easy axis along the chain, these 1D nano-elements turn out
to be quite promising to form bits with a well-defined orientation,
owing to a strong magnetic signal and a large packing density. In
this context, we compute the resonance frequency for both the binding
and anti-binding modes and the resonance field as a function of the
orientation of the applied external magnetic field. This is also motivated
by the fact that the difference in frequency between the two modes
in a pair of ($200$ nm) disks can now be measured with the help of
Magnetic Resonance Force Microscopy as a function of the nanodisks
separation \citep{pigeauetal12prl} of the order $\sim10^{3}$ nm.
Finally, for the coupled chains we found (and calculated) a ``flipping''
magnetic field $h_{f}$ (depending on the shift between the chains
along the major axis) that marks a ``spin-flop'' transition into
a \textcolor{black}{different} magnetic state. This transition could
be of interest in magnetic recording.

\section{System setup and formalism\label{sec:System-setup-and}}

\label{sec:system} 
\begin{figure}
\begin{centering}
\includegraphics[width=0.9\columnwidth]{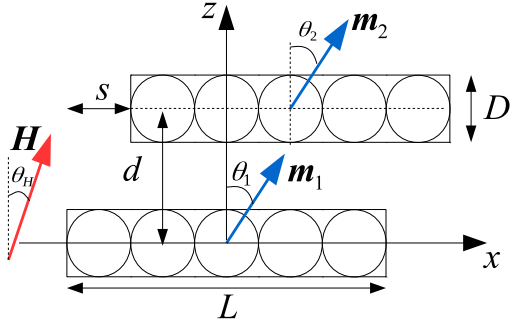}
\par\end{centering}

\protect\caption{\label{fig:TCC}Two coupled chains of identical magnetic nanoparticles
in a transverse magnetic field $\bm{H}$.}

\end{figure}

For the study of the system considered here we
make the approximation that it is composed of nearly parallel chains
of Fe nanoparticles, shifted with respect to each other along their
major axes. Hence, the effects of DI on ferromagnetic resonance in
such assemblies can be studied by first investigating their effects
on a pair of two shifted chains. Each chain is composed of $\mathcal{N}$
identical closely packed nanoparticles. Since these particles are
touching they may be assumed to form a giant magnetic moment with
a strong effective uniaxial anisotropy whose easy axis is along the
chain axis {[}see Section \ref{sub:IsolatedChains}{]}. As such, the
chains can be pictured as cylinders of diameter $D$ and length $L=\mathcal{N}D$,
and diameter $D=2R$. T\textcolor{black}{he system setup consisting
of two chains of nanoparticles, assumed to lie in the $zx$ plane,
is shown in Fig.}\textcolor{blue}{{} \ref{fig:TCC}.}

\subsection{Magnetostatic interaction : beyond the point-dipole approximation}

Our system consists of two chains of nanoparticles, assumed to lie
in the $xz$ plane: chain 1 lies on the $x$-axis and extends from
$-L/2$ to $+L/2$ and chain 2 is parallel to the first chain with
separation $d$ in the $z$-direction and is shifted a distance $s$
along the $x$-axis; it lies from $(-L/2+s,d)$ to $(L/2+s,d)$, see
Fig. \ref{fig:TCC}. 

Inside the chain, the magneto-crystalline anisotropy and shape anisotropy
(for a cylinder with demagnetizing factors $N_{z}=0$, $N_{x}=N_{y}=\frac{1}{2}$)
add up to induce a strong effective anisotropy along the chain's axis
with constant $K_{{\rm eff}}$. The magnetic field $\bm{H}$ is of
variable amplitude and direction and is applied in an arbitrary direction
with respect to the chain's axis. This allows for the calculation
of the resonance frequency as a function of the field amplitude and
the resonance field as a function of the field direction. Consequently,
the energy of a single chain reads (in dimensionless units) 
\begin{equation}
\mathcal{E}\equiv E/\left(K_{{\rm eff}}V\right)=-2\bm{h}\cdot\bm{m}-km_{x}^{2}\label{eq:Energy_free}
\end{equation}
where $h=\mu_{0}HM_{s}/2K_{{\rm eff}}V$. ${\bf M}=M_{s}\bm{m}$,
where $M_{s}$ is the material's saturation magnetization and $\bm{m}$
the unit vector along the (equilibrium) magnetization direction, $V=L\times\pi R^{2}$
is the volume of the cylindrical chain. The symbol $k$ in Eq. (\ref{eq:Energy_free})
is inserted as a flag to identify the anisotropy contribution and
it assumes the value $0$ in the absence of anisotropy and $1$ otherwise.

In addition to the single-chain terms in the energy, chains 1 and
2 are coupled by the DI. We denote by ${\bf M}_{1}=M_{s}\bm{m}_{1}$
and ${\bf M}_{2}=M_{s}\bm{m}_{2}$ the two magnetic moments of the
two chains. In the limit of homogeneous chains with the magnetization
aligned along the chains axis, Escrig \emph{et al.} \citep{escrigetal08apl}
have shown that the interaction energy can be rewritten in the compact
form $E_{\mathrm{int}}=\eta_{\mathrm{int}}\,\bm{m}_{1}\cdot\bm{\mathcal{D}}\bm{m}_{2}$,
where $\bm{\mathcal{D}}$ is the usual dipolar tensor with matrix
elements $\mathcal{D}^{\alpha\beta}=\delta^{\alpha\beta}-3e_{12}^{\alpha}e_{12}^{\beta}$,
where $\delta^{\alpha\beta}$ is the Kronecker symbol and $e_{12}^{\alpha}$
are the Cartesian components of the unit vector ${\bf e}_{12}$ joining
the centers of the chains. Finally, $\eta_{\mathrm{int}}$ is the
coefficient of the DI between two nano-chains as defined in Eq. (4)
of Ref. \onlinecite{escrigetal08apl}.

In the present work, we consider the general case involving anisotropy
and an applied magnetic \textcolor{black}{field} with arbitrary orientation.
Therefore, the magnetic moments ${\bf M}_{1}$ and ${\bf M}_{2}$
may adopt orientations that are not necessarily collinear with each
other and/or with the chains axes. Thus, we derive the dipolar tensor
for this general situation. For this purpose, we regard each chain
as being made up of elementary magnetic moments $d\bm{M}_{i}=\lambda dx_{i}\bm{m}_{i}$,
where $\lambda$ is the linear density of dipoles ($\lambda dx=M_{s}dV$),
$dx_{i}$ the differential element along the chain and $\bm{m}_{i}$
is the unit vector along the magnetic moment of the chain. Since we
are dealing with chains ($L\gg R$), the magnetization $d\bm{M}_{i}$
associated with the differential element $dx_{i}$ can be considered
as being radially uniform. If we use the index $i$ to label these
elements in chain $1$ and $j$ those in chain $2$, the corresponding
elements $ $$d\bm{M}_{i}$ and $d\bm{M}_{j}$ can be considered as
point dipoles interacting via the well-known dipole-dipole interaction
\begin{eqnarray}
dE_{\mathrm{int}} & = & \frac{\mu_{0}}{4\pi}\frac{d\bm{M}_{1}\cdot d\bm{M}_{2}-3\left(d\bm{M}_{1}\cdot{\bf e}_{12}\right)\left(d\bm{M}_{2}\cdot{\bf e}_{12}\right)}{r_{12}^{3}}\label{eq:DDI-energy}\\
 & = & \frac{\mu_{0}}{4\pi}\lambda^{2}dx_{1}dx_{2}\frac{\bm{m}_{1}\cdot\bm{m}_{2}-3\left(\bm{m}_{1}\cdot{\bf e}_{12}\right)\left(\bm{m}_{2}\cdot{\bf e}_{12}\right)}{r_{12}^{3}}\nonumber 
\end{eqnarray}
with

\begin{eqnarray*}
\bm{r}_{12} & = & d\bm{e}_{z}+\left(x_{2}-x_{1}\right)\bm{e}_{x},\quad r_{12}=\left[d^{2}+\left(x_{2}-x_{1}\right)^{2}\right]^{1/2},\\
\bm{e}_{12} & \equiv & \frac{\bm{r}_{12}}{r_{12}}=\frac{d}{r_{12}}\bm{e}_{z}+\frac{\left(x_{2}-x_{1}\right)}{r_{12}}\bm{e}_{x}.
\end{eqnarray*}

Next, upon integrating over the (length of) chains with the corresponding
variables $x_{1},x_{2}$ in the ranges $-\frac{L}{2}\leq x_{1}\leq\frac{L}{2},-\frac{L}{2}+s\leq x_{2}\leq\frac{L}{2}+s$,
we obtain the energy of the DI, taking account of the size and shape
of the chains through the length and the chains separation $d$. More
precisely, this interaction energy can be rewritten for arbitrary
orientations of the two magnetic moments $\bm{m}_{1},\bm{m}_{2}$
as $\mathcal{E}_{\mathrm{int}}=\xi\bm{m}_{1}\cdot\tilde{\mathcal{\bm{D}}}\cdot\bm{m}_{2}$
where 
\begin{equation}
\tilde{\mathcal{\bm{D}}}=\left(\begin{array}{ccc}
\mathcal{I}_{03}-3\mathcal{I}_{25} & 0 & -3d\mathcal{I}_{15}\\
0 & \mathcal{I}_{03} & 0\\
-3d\mathcal{I}_{15} & 0 & \mathcal{I}_{03}-3d^{2}\mathcal{I}_{05}
\end{array}\right)\label{eq:Dipolar-Tensor-gen}
\end{equation}
is the new DI tensor and 
\[
\xi=\frac{1}{K_{{\rm eff}}V}\times\left(\frac{\mu_{0}}{4\pi}\right)\left(\frac{\lambda^{2}}{d}\right)
\]
the new (dimensionless) DI coefficient\textcolor{black}{.} The matrix
elements in Eq. (\ref{eq:Dipolar-Tensor-gen}) are given by the surface
integrals: $\mathcal{I}_{03}=\Theta\left(L,s\right)$, $\mathcal{I}_{05}=2\Theta\left(L,s\right)-\Phi\left(L,s\right)$,
$\mathcal{I}_{15}=\frac{s}{d}\Phi\left(L,s\right)+\frac{L}{d}\left[B^{-1}\left(L,s\right)-B^{-1}\left(L,-s\right)\right]$
and $\mathcal{I}_{25}=\Theta\left(L,s\right)+\Phi\left(L,s\right)$,
with $B\left(L,\pm s\right)=\sqrt{1+\left(L\pm s\right)^{2}/d^{2}}$
and $\Theta\left(L,s\right)=B\left(L,s\right)-2B\left(0,s\right)+B\left(L,-s\right)$,
$\Phi\left(L,s\right)=B^{-1}\left(L,s\right)-2B^{-1}\left(0,s\right)+B^{-1}\left(L,-s\right)$.

Note that $\xi$ is a function of the materials saturation magnetization
$\lambda$ and the chains separation $d$. However, in the results
shown later the DI intensity will be tuned by varying either $d$
or $\xi$ directly. Therefore, the total energy of the system of two
(shifted) chains reads 
\begin{equation}
\mathcal{E}=\sum_{i=1,2}\left(-2\bm{h}\cdot\bm{m}_{i}-km_{x,i}^{2}\right)+\xi\bm{m}_{1}\cdot\tilde{\mathcal{\bm{D}}}\bm{m}_{2}.\label{eq:Total-energy-gen}
\end{equation}

Note that, for convenience, the DI coefficient $\xi$ has been defined
with a dependence on the chains separation as $1/d$. However, the
whole DI term of $\mathcal{E}$ behaves as $1/d^{3}$, as usual, owing
to the dependence on $d$ of the integrals appearing in the matrix
elements of the DI tensor $\tilde{\mathcal{\bm{D}}}$. In addition,
the particular form of these matrix elements is a result of the specific
shape of the elements (here the chains). Therefore, the DI energy
found here with the tensor in Eq. (\ref{eq:Dipolar-Tensor-gen}) take
account of the size and shape of chains, in addition to their separation.

\subsection{Magnetic state of an isolated chain\label{sub:IsolatedChains}}

Before proceeding further we would like to comment on the validity
of the model used here for the magnetic state of the isolated chains
(of nanoparticles) or nanowires. More precisely, we assume that each
chain is a single domain cylinder (or a prolate spheroid) with uniform
magnetization pointing along the chain's axis due to the large shape
anisotropy. The magnetostatic interaction between the two chains is
dealt with using a general approach for computing the demagnetizing
tensor for uniformly magnetized elements of finite-size and arbitrary
shape. This approach extends the so-called point-dipole approximation
which consists in replacing the magnetic elements (here the chains)
by point dipoles. In fact, this assumption fails for finite-size elements
with a too small separation between them and this is why one has to
extend the magnetostatic interaction by including adequate geometrical
factors \citep{beleggiaetal04jmmm272,beleggiaetal04jmmm278,francoetal14jap},
as is done in Eqs. (\ref{eq:Dipolar-Tensor-gen}, \ref{eq:Total-energy-gen}).

There is a huge number of studies on arrays of ferromagnetic nanowires
owing to their promising applications in high-frequency devices. They
have been extensively studied both experimentally and theoretically
\citep{velazquezetal99jap,sampaioetal00prb,thurnetal00sc,demandetal02j3m,lietal04jpcm,zhanetal05prb,nguyenetal06prb,wangetal06nl,climeetal06j3m,piccinetal07epl,kartopuetal09j3m,torreetal09jap,stashkevitch09prb,vidaletal09apl,ott09jap,maureretal09prb,zighemetal11jap,ranall15jap}
with variable length and width. The theoretical work is mainly based
on the numerical approach of micromagnetics or semi-analytical approaches
for solving the extended magnetostatic model. In all of these works,
the magnetization is considered as uniform even in the largest nanowires
(or rather microwires) of an aspect ratio within the range of fabrication
techniques. For example, in Ref. \onlinecite{piccinetal07epl} nanowires
of radius $R$ and length $L$ were studied and even for $R/L$ of
order $10^{-4}$ the magnetization within the nanowires was assumed
to be uniform. It was then shown that the extended model of magnetostatic
interactions reproduces very well the experimental results, see for
instance Fig. 3 of Ref. \onlinecite{piccinetal07epl}. The same conclusion
regarding the validity of this model was reached in other works comparing
theory to the results of other experiments \onlinecite{zhanetal05prb, climeetal06j3m, kouetal09apl}.
In particular, in Refs. \onlinecite{kouetal09apl,boucheretal11apl},
FMR expriments on arrays of nanowires were performed and their results
were favorably compared to the extended magnetostatic model. In these
works, the assembly of nanowires was treated as being organized into
two groups having their uniform magnetization pointing up and down.

\begin{figure}
\begin{centering}
\includegraphics[width=0.95\columnwidth]{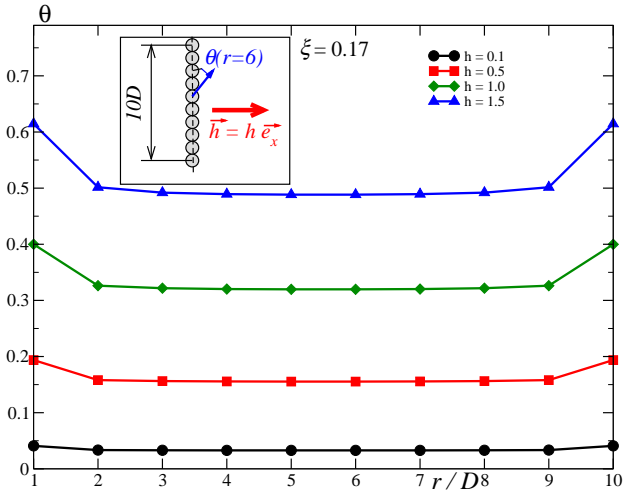}
\par\end{centering}

\protect\caption{\label{fig:Single-chain}Deviation angle $\theta$ of the magnetization
of each nanoparticle as a function of its position in the chain. Inset:
single chain setup consisting of 10 nanoparticles forming a chain
along the $z$-axis with tranverse external field $h$, the deviation
angle $\theta$ is shown in blue for the 6th nanoparticle of the chain.}
\end{figure}

In the present study, the aspect ratio of the chains is $R/L=0.05$.
In addition, the effective anisotropy is dominated by the magnetostatic
(shape) anisotropy with an easy axis along the chain and a constant
$K_{{\rm eff}}\simeq1.6\times10^{6}\,\mathrm{J.m}^{-3}$. These specifications
put on the safe side the assumption of uniform magnetization with
an easy direction along the chain's axis. In other words, the chains
are magnetically saturated along their axes. This is obviously more
so in the case of an external magnetic field applied along the chains.
In the present study the direction of the magnetic field is varied
with respect to the chains axes. In order to ensure that the assumption
of uniform magnetization still applies even in the most unfavorable
situation of a field applied perpendicular to the chains axis, we
have performed numerical calculations for (isolated) chains of 10
spherical nanoparticles along the $z$-axis, as shown in the inset
of Fig. \ref{fig:Single-chain}. The nanoparticles constituting each
chain have a diameter $D=20\ {\rm nm}$ and an (effective) easy anisotropy
axis along the chain. Within the chain the nanoparticles interact
with each other via the long-range DI with strength $\xi_{{\rm intra}}=\left(\frac{\mu_{0}}{4\pi}\right)\frac{\pi}{6}M_{s}^{2}/K_{{\rm eff}}$,
which evaluates to $\xi_{{\rm intra}}\simeq0.17$ for iron \textcolor{black}{(}$M_{s}=1.7~10^{6}\ {\rm A.m^{-1}}$).
In Fig. \ref{fig:Single-chain} we plot the deviation angle of the
individual magnetic moments of the nanoparticles within the chain
as a function of their position in the chain. 

It is clearly seen that even in a transverse magnetic field the magnetic
moments within the chain tilt towards the field direction in unison
apart from a small number of them located at and near the chain's
ends. However, the relative deviation of these boundary moments is
rather small. Similar results were obtained from micromagnetic calculations
in Ref. \onlinecite{zighemetal11jap}. This result is due to the fact
that, within the chain, the dipolar interactions favor the (super)ferromagnetic
state as they induce an extra anisotropy along the chain. More precisely,
they renormalize the anisotropy as $k\to k_{{\rm eff,DI}}$. \textcolor{black}{The
idea underlying this renormalization can be illustrated by considering
a chain of $\mathcal{N}$ free nanoparticles with a uniaxial anisotropy
$k$ and a transverse external field $h$, in the absence of DI. The
deviation angle at equilibrium is then given by $\sin\theta_{{\rm eq,Free}}=h/k$.
For low-to-intermediate magnetic fields, the deviation angle remains
small, such that $\sin\theta_{{\rm eq,Free}}\simeq\theta_{{\rm eq,Free}}$.
In this case, after switching on the DI the deviation angle can be
self-consistently determined to first order in $\xi_{{\rm intra}}$.
Indeed, it can be shown that it is given by $\theta\left(r_{i}\right)=h/k_{{\rm eff,DI}}\left(r_{i}\right)$,
where $k_{{\rm eff,DI}}$ depends only weakly on the position $r_{i}$
within the chain since $k_{{\rm eff,DI}}\simeq k\left[1+3\xi_{{\rm intra}}I\left(r_{i}\right)\right]$.
The lattice sum $I\left(r_{i}\right)$ stems from the intra-chain
DI between a particle and all the other ones within the chain (}\textcolor{black}{\emph{i.e.}}\textcolor{black}{{}
those on its left and its right), namely
\begin{equation}
I\left(r_{i}\right)=\sum_{j=1}^{i-1}\frac{1}{\left(r_{i}-r_{j}\right)^{3}}+\sum_{j=i+1}^{\mathcal{N}}\frac{1}{\left(r_{j}-r_{i}\right)^{3}}.\label{eq:lattice-sum}
\end{equation}
The symmetry of the expression above translates into the symmetry
of the deviation angle depicted in Fig. \ref{fig:Single-chain}. }We
note in passing that for a chain of 10 spherical nanoparticles, we
obtain the effective anisotropy $k_{{\rm eff,DI}}\simeq2k$, which
exceeds the largest value of the magnetic field used in Fig. \ref{fig:Single-chain}.

To sum up, these additional calculations do confirm that the magnetization
within an (isolated) chain may be reasonably regarded as uniform even
in a transverse magnetic field. On the other hand, for the magnetostatic
interaction between the two chains, we have developed an extended
model with a new magnetostatic tensor {[}see Eq. (\ref{eq:Dipolar-Tensor-gen}){]}
that takes into account the finite size and shape of the interacting
nano-elements.

In the following we present and discuss our results for the system
of two coupled chains as described above.

\section{Results\label{sec:Results}}

\subsection{FMR characteristics \label{sub:FMR-characteristics}}

\textcolor{black}{The FMR characteristics, namely}\textcolor{black}{\emph{
}}\textcolor{black}{the resonance frequency and resonance field, are
computed as follows. For a given system configuration (including anisotropy,
applied field and spatial configuration of the two chain), we first
determine the equilibrium state of the system, }\textcolor{black}{\emph{i.e.}}\textcolor{black}{{}
the spatial orientation of the two (macroscopic) magnetic moments
of the chains. Then, we linearize the Landau-Lifshitz equation around
this state leading to an eigenvalue problem for the system. Upon solving
the latter for a given applied magnetic field (with given amplitude
and direction) we obtain the various eigenfrequencies of the excitation
modes of the coupled two chains. Next, for a fixed frequency we solve
the eigenvalue problem for the applied magnetic field and this renders
the resonance field.}

Now, we proceed to compute the resonance frequency $\omega_{{\rm res}}$
and the flipping field $h_{f}$ of the system whose total energy is
given in Eq. (\ref{eq:Total-energy-gen}). We first compute the resonance
frequency for the case of zero shift ($s=0$) between the two chains
as this leads to tractable analytical expressions. For the general
case, $\omega_{{\rm res}}$ and $h_{f}$ will be computed numerically,
respectively as a function of the field amplitude $h$ and the field
direction $\left(\theta_{h},\varphi_{h}\right)$. 

For the non-shifted chains, the analytical calculation of $\omega_{{\rm res}}\left(h\right)$
is done for the setup with $\theta_{h}=0,\varphi_{h}=0$. In the present
case, the two anisotropy axes and the applied magnetic field are all
in the $zx$ plane, and as such the magnetic moments also lie in the
same plane, \emph{i.e.} we have $\bm{m}_{i}\left(\theta_{i},\varphi_{i}=0\right)$.

In the absence of anisotropy and applied field, the DI favors a ferromagnetic
order of the two magnetic moments along the dimer's bond, \emph{i.e.}
along the $z$ axis. If the effective uniaxial anisotropy is added
with easy axis along the chains, the two magnetic moments order anti-ferromagnetically
along the $x$ axis. Finally, when the magnetic field is applied along
the $z$ axis, the two magnetic moments are tilted to an oblique angle
that depends on ($h,\xi$), \emph{i.e.} a canted anti-ferromagnetic
state. Upon analyzing the energy stationary points, it turns out that
there are two field regimes separated by the critical value $h_{c}=k\left[1-\tilde{\xi}\left(\delta^{2}+a\right)\right]$,
where $a=1-\sqrt{1+\delta^{2}}$ and $\delta=L/d$ and $\tilde{\xi}=\xi/\left(k\sqrt{1+\delta^{2}}\right)$.
More precisely, the polar angles of the two magnetic moments are $\theta_{1}=-\theta_{2}=\theta^{\left(\mathrm{m}\right)}$
with $\cos\theta^{\left(\mathrm{m}\right)}=h/h_{c}$ for $h\leq h_{c}$
(and $\theta^{{\rm (m)}}=0$ otherwise). This result obviously coincides
with that obtained in Ref. \onlinecite{francoetal14jap}, Eq. (37)
after a rotation of the frame axes and noting that for point dipoles\emph{
$\delta=L/d$} becomes small and that the quantity $\left(\frac{\mu_{0}}{4\pi}\right)\left(\lambda L\right)^{2}/d^{3}$
is the DI coefficient in Ref. \onlinecite{francoetal14jap}.

For $h\leq h_{c}$, we obtain the following analytical expressions
for the resonance frequencies of the binding (B) and anti-binding
(AB) modes,

\textcolor{black}{
\begin{eqnarray}
{\normalcolor {\color{blue}{\normalcolor \tilde{\omega}_{\mathrm{B}}}}} & = & \left(2k\right)\sqrt{1-\left(h/h_{c}\right)^{2}}\times\sqrt{\left(h_{c}/k\right)\left[1-\left(2a+\delta^{2}\right)\tilde{\xi}\right]},\nonumber \\
{\color{blue}{\normalcolor \tilde{\omega}_{\mathrm{AB}}}} & = & \left(2k\right)\sqrt{\left[1+\left(\delta^{2}-a\right)\tilde{\xi}\right]-\left(h^{2}/h_{c}^{2}\right)\left[1+\left(\delta^{2}+a\right)\tilde{\xi}\right]}\nonumber \\
 &  & \times\sqrt{1+\delta^{2}\tilde{\xi}}.\label{eq:OmegasNoShiftLowField}
\end{eqnarray}
}

\textcolor{black}{Here $\tilde{\omega}$ is the dimensionless frequency
defined by $\tilde{\omega}\equiv\omega/\omega_{{\rm a}}$, where $\omega_{{\rm a}}=\gamma H_{a}$
with $H_{a}$ being the anisotropy field given by $H_{a}=2K_{{\rm eff}}/M_{s}$.
For the material parameters given earlier, our reference frequency
is then $\nu_{a}=\omega_{{\rm a}}/2\pi\simeq52\,{\rm GHz}$. }

\begin{figure}
\begin{centering}
\includegraphics[scale=0.3]{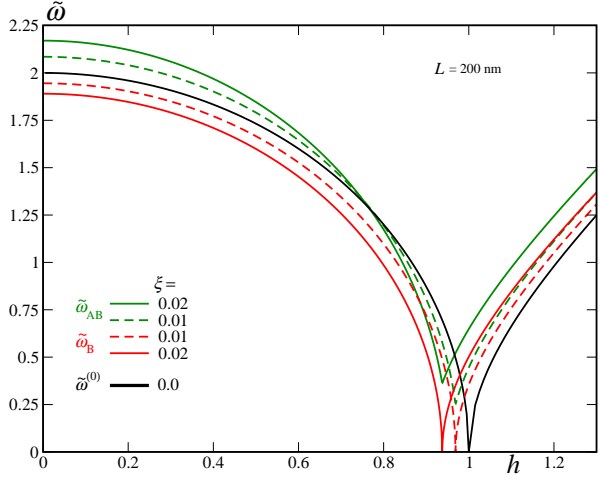}
\par\end{centering}

\protect\caption{\label{fig:o1o2s=00003D00003D00003D0-h}Resonance frequencies of the
binding (red) and anti-binding (green) modes for non shifted chains,
for different \textcolor{blue}{${\normalcolor \xi}$}. }
\end{figure}

\textcolor{black}{For non interacting chains one recovers the well
known result of two degenerate modes with the frequency $\tilde{\omega}_{\mathrm{B}}=\tilde{\omega}_{\mathrm{AB}}=\left(2k\right)\sqrt{1-\left(h/k\right)^{2}}\equiv\tilde{\omega}^{\left(0\right)}$}.
By inspection of Eq. (\ref{eq:OmegasNoShiftLowField}), we see that
for small $\xi$ and $h$ the frequency of the anti-binding mode is
higher than that of the binding mode because it corresponds to an
out-of-phase precession of the magnetic moments as seen in Fig. \ref{fig:o1o2s=00003D00003D00003D0-h}.

For $h>h_{c},$ the resonance frequencies are given by
\begin{eqnarray}
\tilde{\omega}'_{\mathrm{B}} & = & 2h\sqrt{\left(1-\frac{a\tilde{\xi}}{h}\right)}\sqrt{1-h_{c}/h},\label{eq:OmegasNoShiftHighField}\\
\nonumber \\
\tilde{\omega}'_{\mathrm{AB}} & = & 2h\sqrt{1+\frac{\left(a+2\delta^{2}\right)\tilde{\xi}}{h}}\sqrt{1-\frac{h_{c}+2a\tilde{\xi}}{h}}.\nonumber 
\end{eqnarray}

\textcolor{black}{In Fig. \ref{fig:o1o2s=00003D00003D00003D0-h} we
observe a shift (downwards) of the critical field $h_{c}$ at which
(only) the binding modes vanishes as the intensity of the DI increases.
}This is obviously recovered by the expression $h_{c}$ of the critical
field given above. The reason for this effect is that the energy minimum,
with $\cos\theta^{\left(\mathrm{m}\right)}=h/h_{c}$, is a result
of the competition between the anisotropy and the combined effect
of the applied field and the DI. Thus, when the latter becomes stronger,
a weaker field is needed to overcome the effect of the anisotropy.

\begin{figure}
\begin{centering}
\includegraphics[scale=0.3]{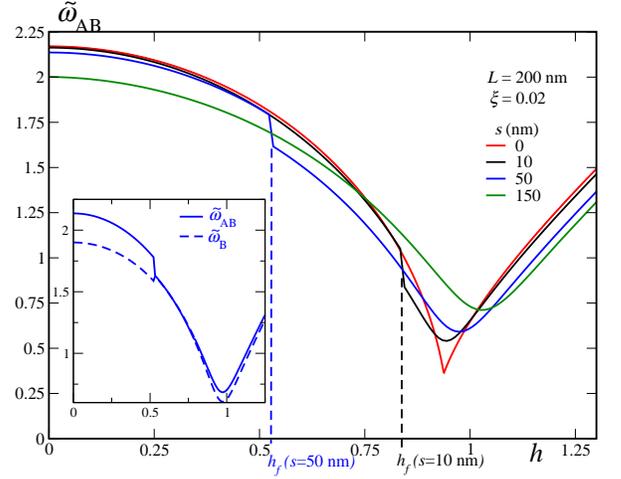}
\par\end{centering}

\protect\caption{\label{fig:OmegaB-Shift}Resonance frequency $\tilde{\omega}_{{\rm AB}}$
for shifted chains. \textcolor{black}{The inset shows the binding
and anti-binding modes for}\textcolor{blue}{{} ${\normalcolor s=50\ {\rm nm.}}$}}
\end{figure}

In Fig. \ref{fig:OmegaB-Shift} we show the effect on the frequency
of the \textcolor{black}{anti-binding} mode of a shift of the chains
with respect to each other along their axes. 
 These results show that for a given shift between the two chains,
there appears a jump in the resonance frequency of the \textcolor{black}{anti-binding}
mode - that of\textcolor{blue}{{} }\textcolor{black}{the binding mode
also exhibits such a jump (but less pronounced) for the same value
of the $h_{f}$ (Inset in Fig. \ref{fig:OmegaB-Shift})}. This jump
can be understood as follows. For very small fields, the equilibrium
state is an anti-ferromagnetic ordering of the two magnetic moments
along the anisotropy axes. As the field is increased, there is a transition
into a ferromagnetic state in an oblique direction with respect to
the applied field. It is this transition that is responsible for the
abrupt change in the resonance frequency. A further increase of the
magnetic field leads to the saturation state and thereby to the asymptote
in the form of a straight line ($h\ge h_{c}$). As the shift of the
two chains increases, the field at which this jump occurs (we call
it the spin-flop field $h_{f}$) decreases. Indeed, as the shift is
increased, the dimer's bond tilts towards the chain axes and thereby
the two magnetic moments tend to order along the anisotropy easy axes.
In this case the ferromagnetic order is more favorable and this is
why the field amplitude required to trigger the transition from the
anti-ferromagnetic to ferromagnetic order is smaller. The system then
rotates towards the direction of the applied field as a single (larger)
magnetic moment. For a large shift, the system is in a ferromagnetic
order already at zero field, see the green curve in Fig. \ref{fig:OmegaB-Shift}.

\begin{figure}
\begin{centering}
\includegraphics[scale=0.3]{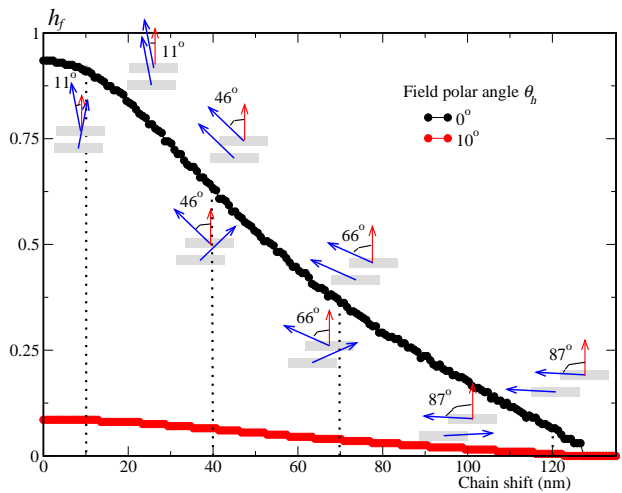}
\par\end{centering}

\protect\caption{\label{fig:hcVsS-thetah}Flipping magnetic field $h_{f}$ marking
the ``spin flop'' transition as a function of the chains shift,
for two field directions\textcolor{black}{{} and for}\textcolor{blue}{{}
${\normalcolor \xi=0.02}$}. For $\theta_{h}=0\text{\ensuremath{\protect\textdegree}}$,
the magnetic moments are drawn in blue for various shifts below and
above the ``spin-flop'' transition. }
\end{figure}

For a magnetic field applied in the $z$ direction ($\theta_{h}=0$),
the black curve in Fig. \ref{fig:hcVsS-thetah} clearly illustrates
the decrease of the field $h_{f}$ as a function of the chain shift
$s$. As soon as the applied magnetic field is tilted with respect
to the $z$ direction ($\theta_{h}\neq0$), the field $h_{f}$ at
\textcolor{black}{which the transition between the two ordered states
occurs reduces significantly.} 
 This evolution is clearly seen in Fig. \ref{fig:hcVsS-thetah}, where
we compare the evolution of the direction of the two magnetic moments
for $\theta_{h}=0\text{\ensuremath{\textdegree}}$ and $10\text{\ensuremath{\textdegree}}$.
First, as we go vertically through the $h_{f}$ curve, at a given
shift $s$, we observe the switching of one of the two magnetic moments:
the $x$-component of the moments switches from an anti-ferromagnetic
to a ferromagnetic order. Second, as we increase $s$ we see that
the canting angle of the ``anti-ferromagnetic'' order, in the phase
with $h<h_{f}$, becomes larger ending up in a nearly complete anti-ferromagnetic
order along the anisotropy axis. Finally, as discussed above, beyond
some critical value of the shift, that depends on $L$ and $d$, there
is only one phase corresponding to the ferromagnetic order. The maximum
critical shift $s_{0}$ corresponding to $h_{f}=0$ (\emph{e.g.} in
Fig. \ref{fig:hcVsS-thetah} $s_{0}\sim129{\rm nm}$) can easily be
predicted as a function of $L$ and $d$. Indeed, if one considers
ferromagnetic and anti-ferromagnetic states fully polarized along
$x$ on either side of this point, the DI energies of these states
should be the same at the transition. Hence, $s_{0}$ is the root
of the equation $\Phi\left(L,s_{0}\right)=0$, which implies that
the interaction coefficient $\eta_{{\rm int}}$ vanishes (\emph{i.e.}
DI$\to0$). As we intuitively expect, $s_{0}$ increases as $d$ or
$L$ increases. 

\begin{figure}
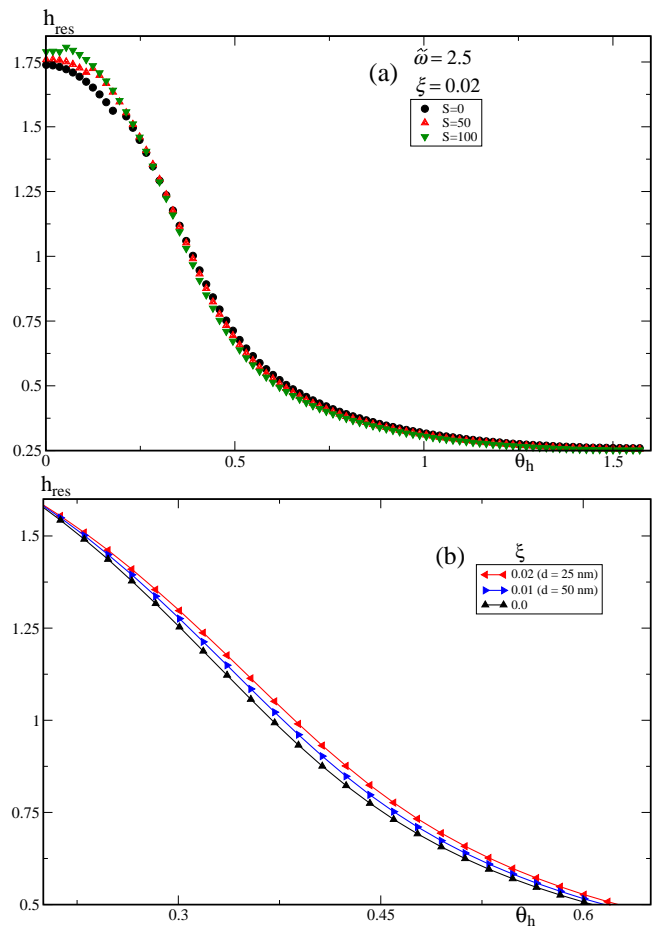

\begin{centering}
\includegraphics[scale=0.3]{figure6a}
\par\end{centering}

\begin{centering}
\includegraphics[scale=0.29]{figure6b}
\par\end{centering}

\protect\caption{\label{fig:hres}Resonance field against the applied field polar angle
$\theta_{\mathrm{h}}/\pi$, for variable spatial shift\textcolor{black}{{}
(a) and inter-chains separation with $s=0$ (b).} }
\end{figure}

Finally, in FMR experiments one routinely obtains the resonance field
as a function of the direction of the applied magnetic field. The
corresponding data is an efficient means for characterizing the system
with regards to the easy/hard magnetization directions. So it is worthwhile
to compute this observable for our system and to investigate the effect
of a chains relative shift. Accordingly, we have (numerically) computed
the resonance field as a function of the applied field polar angle
$\theta_{h}$ upon varying the spatial shift and inter-chains separation
(or DI strength). The results are shown in Fig. \ref{fig:hres}. 
 It can be seen that the resonance field exhibits the usual overall
behavior as a function of the applied field direction, as the latter
rotates from the anisotropy easy axis to the hard axis \citep{gurmel96crcpress}.
On the other hand, in the present study the resonance field shows
a weak dependence on the spatial shift $s$, whereas its dependence
on the inter-chains separation $d$ follows the expected behavior.
More precisely, for small angles $\theta_{h}$ the applied field competes
with the DI and this leads to higher resonance fields for higher DI
strengths (or smaller separation $d$).

\section{Conclusions}

We have computed the FMR characteristics of a system of two coupled
chains, taking into account their separation and relative shift. The
DI has been dealt with taking into consideration their finite size
and shape. We have found that the shift of the two chains along their
axes has a significant effect on the resonance frequency. More precisely,
as the magnitude of the magnetic field is increased the system goes
through a ``spin-flop'' transition from an anti-ferromagnetic order
to a ferromagnetic order before reaching the high-field branch of
the resonance frequency. The field that marks this transition decreases
with an increasing shift of the chains and depends on the systems
specifications such as the length of the chains, their separation
and the orientation of the magnetic field. This field defines magnetic
regions of importance for magnetic recording media made of 1D nanoelements. 

We have also computed the resonance field as a function of the magnetic
field direction for varying inter-chain separation and spatial shift.
The resonance field is what is routinely measured in FMR measurements
with a rotating applied magnetic field and allows for characterizing
the system with regards to its physical parameters. In the present
study, this could be useful for characterizing, \emph{inter alia},
the magnetostatic interaction between the chains.

Finally, the present study, restricted to a dimer, allows one to fully
investigate the critical shift as a function of the applied field
(which mimics the write/read process), and sets the stage for further
investigation involving dimer assemblies. The latter could in principle
be tackled numerically using the present approach by summing over
pairs with the effective DI derived here.
\begin{acknowledgments}
We acknowledge a useful discussion with David Schmool.
\end{acknowledgments}

\bibliographystyle{apsrev}
\bibliography{hkbib}

\end{document}